\documentstyle[12pt] {article}
\begin{document}

{\center{\large{\bf GRB 011211: An alternative interpretation of the 
optical and X-ray spectra in terms of blueshifts\\}}}

{\bf D. Basu}\footnote{Department of Physics, Carleton University, 
Ottawa, ON K1S 5B6, Canada. e-mail: basu@physics.carleton.ca\\}

{\bf Abstract} The redshift of the gamma ray burst (GRB) GRB 011211 has 
been determined
as 2.14 from several absorption lines seen in the spectrum of its optical
afterglow. The spectrum of its X-ray afterglow exhibited several emission 
lines, and their identification 
led to a mean
redshift 1.862. A supernova model has been proposed based on the
redshift 
of the GRB as 2.141. It is shown here that the redshift
interpretation cannot explain the observed spectra, as  
some serious 
inconsistencies exist in the process of redshift
determinations in spectra of both optical and X-ray afterglows. In view of
that, an
alternative interpretation of the spectra is presented in terms of
blueshifts. Ejection mechanism is proposed as a possible scenario to 
explain the blueshifted spectrum.\\

PACS Nos.: 98.38.Mz: 98.62Py; 98.80.-K
\pagebreak

\def\baselinestretch{1.66}
\large
\normalsize

{\bf 1. Introduction\\}

Although the first gamma ray burst (GRB) was discovered over thirty years
ago [1], it is only in recent years that redshifts are being
determined after their optical counterparts are associated with host
galaxies. As a result, GRBs are considered as extragalactic objects due to
their high redshift values. However, high redshifts imply enormous
distances and this has raised the problem of energetics. Meanwhile, 
several models
have been proposed to account for the extremely high energies required,
and
involve merging of neutron stars (or black holes) in a binary [2], or 
collapse of massive stars producing black holes with superstrong
($\approx$10$^{15}$ G) magnetic fields (supernova (SN)/hypernova (HN) 
models [3,4].

GRB 011211 was detected [5] by the BEPPO-SAX
satellite on 11 December 2001 and its redshift was determined [6,7]
from several absorption lines seen in the spectrum of its optical
afterglow as z$_{r(op, abs)}$ = 2.14. A new 
absorption spectrum of the object with S/N superior to that of Holland et 
al. [7] has recently
been published [8], and this is quite different from 
the earlier one with only four common features. Also, as Vreeswijk et al. 
[8] points out, some of the lines in Holland et al. [7] were 
actually misidentified. The mean absorption redshift obtained by Vreeswijk 
et al. [8] is 2.1418 based on the identification of sixteen out of the
seventeen lines seen in the spectrum and originating in a single 
absorption system.

GRB011211 was subsequently observed by the XMM-Newton
X-ray telescope and
the spectrum of the X-ray afterglow exhibited "an apparent
absorption feature" and five emission lines [9]. The former was
identified on the assumption that it "arises in the same material as the
line emission", while the
identification of
the five emission lines with highly ionized metal lines led to a mean
redshift z$_{r(x,em)}$ = 1.862. Reeves et al. [10] have, of course, 
re-analysed their data, and have withdrawn the absorption line which is 
not real. 

Nevertheless, there appears to be some controversy over the analysis of 
the 
X-ray spectrum by Reeves et al. [9,10]. Borozdin \& Trudolyubov 
[11] analyzed the first 5ksec of the Reeves et al. [9,10] data 
for both PN alone and combined PN, MOS1 and MOS2. A good fit for an 
absorbed power law with Galactic absorption was found for the combined 
data showing no improvement from adding lines. But for the PN data 
alone, which the Reeves et al. [9,10] analysis is based on, 
improvement to 
the fit at 99.9 percent confidence level was found when lines identified 
as partially ionized 
lines 
of Mg, Si, S, Ar, Ca were added with a mean redshift of 1.9. On the other 
hand, independant analyses by Rutledge \& Sako [12] and Sako et al. 
[13] suggested the statistical significance of the features to be 
marginal.

But Butler et al. [14] demonstrated that the 
conflicting estimation of the statistical significance by Reeves et 
al. [9,10], and by Rutledge and Sako [12] and Sako et al. [13], 
has  
arisen from different assumptions in the continuum modelling. Both methods 
would agree that the lines are statistically significant at 
$\approx$3$\sigma$ level, as originally claimed by Reeves et 
al. [9], if the column density is taken as the Galactic value as in the 
analysis of Reeves et al. [9]. Butler et al. [14] have further shown 
that 
the density parameter should actually be taken as the Galactic value for 
the first 5ksec portion of the XMM-Newton EPIC-PN spectrum where the 
potential emission lines are located, since the assumption of the Galactic 
absorption leads to well modelling of the full 25ksec data. This 
confirms that the emission lines reported by 
Reeves et al. [9,10] {\it are} real and statistically significant.

The importance of the correct estimation of redshifts in 
building models of extragalactic objects, including GRBs, can hardly be 
underestimated, as redshifts are essential for determining the energetics 
of the objects. 
The model proposed [9,10] for the GRB
011211 is an SN model, the computed total equivalent isotropic energy of
5x10$^{52}$ erg being based on the redshift 2.141. 

One has to keep in mind that all line identification processes are  
programmed, almost as a rule, to 
determine 
redshifts only, by identifying search lines to observed lines at the red 
side. No attempt is made to identify search lines with observed lines 
at the blue side, and, as such, blueshifts are not considered. Possibility 
of misidentification of observed lines, and, hence, of blueshifts in 
spectral lines of extragalactic objects have, of course, been pointed out 
by several researchers in 
the past [15-17] Observed 
lines are indeed sometimes misidentified for redshift determinations. 
Spectra of 
objects like GRB 011211 that cannot be explained in terms of redshifts, 
should be particularly considered and re-examined in this respect. The 
purpose of this paper is to show that an
alternartive interpretation in terms of blueshifts can explain the
spectra.

Sec. 2 reviews some specific examples in the literature to 
demonstrate misidentifications of observed lines and their possible 
blueshift (z$_{b}$)
interpretations. Serious inconsistencies in the redshift (z$_{r}$) 
deteremination in 
the spectra of both optical and x-ray afterglows of GRB 011211 are 
demonstrated 
in Sec. 3, and the spectra is interpreted in terms of the alternative 
blueshift hypothesis in Sec. 4. Ejection mechanism is proposed 
as a possible scenario 
to explain the blueshifted spectra in Sec. 5. and some concluding remarks 
are presented in Sec. 6.

{\bf 2. Examples of misidentification of lines and possible 
blueshifts\\}

It is known that CIV 1549 and CIII] 1909 are two of the strongest search 
lines in the redshift identification system, and if one of them is seen 
the 
other should be seen as well in the observed spectrum [18]. However, 
published spectra in 
the high redshift galaxy (HRG) 0316-257B (z$_{r}$=3.1351) exhibits CIV, 
but there is no line at 7894\AA, the expected position of CIII]  
[19], and the authors do not even mention about the CIII]. Same is the 
case with another HRG Obj 19 (z$_{r}$=2.39) which 
exhibits CIV [20], and a very 
insignificant and most like likely noise signal is claimed as the CIII] 
at the expected wavelength of 6472\AA, marked doubtful. 
The 
spectra 
of these two and another 13 HRGs have 
been explained by re-identifications of observed lines in terms of 
blueshifts [21].

Again, slitless spectroscopy of the galaxy STIS123627+621755 
demonstrates [22]  an emission line around 9334\AA, followed by a 
discontinuity starting around 9300\AA. A z$_{r}$=6.68 was determined by 
the identification of the line with Ly$\alpha$ 1216 and the discontinuity 
as the Ly decrement. However, subsequent observation at B,V 
bands 
[23] found the identifications wrong and concluded that "the redshift is 
undetermined", as Ly$\alpha$ limit (912\AA) is shifted to 7004\AA at 
z$_{r}$=6.68 and no flux should be observed below 7004\AA, i.e. at B,V 
bands. The same galaxy was also observed at 6700\AA and a 
non-detection was reported at 1.2 micron [24] which also ruled out the 
z$_{r}$ of 6.68. The spectra has been successfully interpreted as 
blueshifted [25].

Furthermore, the spectrum of the QSO PG 1407+265 is very unusual in 
the 
sense that the major UV lines viz. Ly$\alpha$, CIV, CIII, MGII 2798, are 
very weak and H$\alpha$ is also much weaker than normal, 
while the usually weak FeII forest [26] are "unusually strong" [27]. 
Several 
attempts to explain the spectrum turned unconvincing and the nature of the 
spectrum "remains puzzling" [27]. 

In addition, the spectrum of PG 1407+265 also 
exhibits a large number of absorption lines leading to the determination 
of several absorption redshift systems [28]. However, the 
identifications of the absorption lines show many inconsistencies. 
Some lines remain unidentified, several identifications are reported 
doubtful, the same search line (Ly$\alpha$) has been identified with more 
than one observed lines for the same redshift system in many cases, 
several redshifts have ben computed by identification of a single line, 
'spread' values are larger than the usually accepted value of 0.01 for 
absorption redshift syatems in most cases, a redshift system has been 
computed on the basis of identification of a higher order Lyman line with 
a 
stronger observed line and a lower order Lyman line with a weaker observed 
line.
All these discrepancies and inconsistencies lead to the 
absorption redshift systems unacceptable.

PG 1407+265 and unusual spectra of two other QSOs, viz. SDSS 1533-00 and 
PKS 0637-752 have been explained on 
the basis of blueshift hypothesis [29]. Additionally, observed spectra of 
25 other QSOs available in the published literature have also been 
identified 
with search lines of longer wavelengths and blueshifts determined [30]. 
The spectra of another QSO (radio loud), viz. PKS 2149-306, which could 
not be explained 
in the usual redshift interpretation has been successfully 
re-interpreted as blueshifted [31].

Moreover, QSO pairs seen {\it across} active galaxies are believed to be 
ejected  from the galaxy involved. It is, however, more logical that the 
pair 
should be ejected in opposite directions with equal probability, and, as 
such, one should be moving away from us exhibiting redshift, while the 
other should be approaching us exhibiting blueshift, rather than 
both exhibiting redshifts implying both are ejected away from the 
observer. Analysis of four pairs demonstrated that the observed spectrum 
of 
one object in each pair can be interpreted as blueshifted [32]. 

Further, the redshift of the host galaxy of the GRB 971214 has been 
eveluated by the identification of one emission line and "a drop  
immediately on the blue (short wavelength) side of the line", in addition 
to the 
absence of any flux "blueward of 4030\AA the redshifted Ly$\alpha$ 
continuum break" [33]. However, the published record 
shows that 
the 'drop' is not acceptable as its magnitude is of the same order as the 
noise level. On 
the other hand, the record stops at 4000\AA which makes the argument of 
"no flux blueward of 4030\AA" unconvincing. The spectra the host galaxy of 
971214 and 
three other GRB host galaxies have been interpreted as blueshifted 
[34].

Again, the observed spectrum of the host galaxy of the SN Ia 96T 
exhibits three emission features identified in the redshift scenario as 
H$\alpha$, [OIII] 5007 and [OII] 3727 [35]. However, examination of 
the profiles of the lines revealed that the lines at 8141\AA and 6212\AA 
identified with H$\alpha$ and [OIII] 5007 respectively are unacceptably 
week for the 
two recognized strong search lines [30], actually 
weaker 
than the line at 4626\AA, identified with the [OII] 3727 (unfortunately, 
no 
equivalent widths are 
avilable). The spectra of the host galaxy of 96T and host galaxies of four 
other SNe Ia were re-interpreted as blueshifted, and blueshifts 
determined 
by re-identification of the observed lines with search lines of longer 
wavelengths [36].

Finally, the puzzling spectrum of the galactic 
X-ray 
source 1E 1207.4-5209 has been explained as blueshifted and shown to be 
due to two ejected absorbing clouds originating at the centre of the SNR 
G296.5+10.0 [37].

{\bf 3. Inconsistencies in the redshift determination of GRB 011211\\}

3.1. Optical absorption lines\\

The assumption that the redshift of
the GRB is equivalent to z$_{r(op,abs)}$ = 2.141 implies
that {\it all} the absorption
lines in the optical spectrum arise in the host galaxy [10]. In reality, 
this
is the lower limit of the GRB redshift. The {\it redshift} of an
extragalactic object is the {\it emission redshift} determined from the
{\it emission lines}.
Radiation from the object is more likely to encounter several 
absorbing
clouds in the
line of sight between the object itself and the
observer. Redshifts determined from absorption lines arising in these
clouds, i.e. absorption
redshifts, are therefore, in general, smaller than the redshift of
the object if redshifts are cosmological. The number of absorption
redshift systems for an extragalactic object with the redshift 
around 2.2 may be as high as 20 [38]. However, some lines may
arise in the host galaxy, in which case the largest of the absorption
redshifts will be equal to the redshift of the host galaxy, the latter
being determined, once again, from emission lines. Hence, more than one 
absorption
redshift systems, all smaller than (although one may be equal to) the
emission redshift, is expected. Spectra with different  
features, viz. emission and absorption,  have been reported for several 
other GRBs, as discussed below, which would 
support the scenario presented above.

GRB 021004 exhibits many absorption lines that have been identified 
with search lines leading to as many as {\it five} absorption systems , 
viz. 1.3806, 
1.6039, 2.2983, 2.3230 and 2.3293, while the single emission line 
identified with Ly$\alpha$ 1216 yields the emission redshift 2.3351 
[39]. Thus, the largest absorption system is nearly equal to the 
emission 
system, i.e. the redshift of the host galaxy, with a difference of only 
0.0058. This makes the largest system arising in the host galaxy 
itself, and the other absorption systems, smaller than the emission 
system, arising in the intervening absorbing clouds under cosmological 
hypothesis [39]. These redshifts are also confirmed by 
Mirabal et al. [40], who equate the absorption redshift 2.328 of the AlII 
1670.71 
line to the Ly$\alpha$ 1216 emission redshift 2.328, computed from their 
data. 
and concludes that this 
highest system originates in the host galaxy. 

Another object, viz. GRB 020405, was detected  at radio and X-ray 
wavelengths [41,42], although it did 
not 
exhibit any discrete feature in its X-ray spectrum, neither emission nor 
absorption. This has been interpreted as the effect of a "long lasting 
bright afterglow" that might have been responsible for the non-detection 
of any faint discrete feature [41]. Nonetheless, the 
optical spectrum of the host galaxy exhibited Balmer and oxygen emission 
lines 
which yielded the redshift 0.691 [43]. Additionally, 
the optical spectrum is also rich in absorption lines, showing  
twelve features which have been identified with FeII and MgII 
lines, yielding {\it two} absorption systems, viz. 0.691 and 0.472. Once 
again, the 
former, the larger of the two, being equal to the emission system, 
originates in the 
host galaxy itself, and the latter is an intervening system identified by 
the imaging technique with 
a cloud in the galaxy complex [43].

It thus appears that spectra of GRB host galaxies, rich in absorption 
lines and also exhibiting emission features, is not uncommon. In such 
cases, the 
absorption lines are interpreted as {\it multiple} absorption redshift 
systems with 
the 
largest, if and when having a similar value as the emission redshift 
system, 
originating in the host galaxy, and others, having smaller values, 
originating in intervening space. Apart from the papers quoted above, 
reference can also be made to [44,45], where 
presence of intervening systems have been 
reported for the GRB host galaxy spectra.

In case of the present object, viz. GRB 
011211, the assumption that the redshift of the
object is equal
to a single absorption redshift exhibited by {\it all} the absorption 
lines, viz. as many as seventeen, is an
oversimplification of the situation. Therefore, even if
the redshift scenario is
correct, the
redshift of the GRB 011211 is
most likely much larger than 2.141. This puts the GRB at a much
larger
distance,
and hence involves much larger energy than the energy of a
typical supernova, viz. 5x10$^{52}$ erg, which the model is based on.  

The importance of the above arguments is further evident in the serious 
inconsistency in the determination of the absorption redshift 
z$_{r(op,abs)}$ = 2.1418 [8], which cannot identify 
{\it 
all} the lines in the spectrum, the line at 6114.2\AA remaining 
unidentified. This clearly shows that a single system has failed to 
explain the observed absorption spectrum in the redshift interpretation. A 
redshift system cannot be accepted unless {\it all} the lines in the 
system are identified exhibiting the {\it same} redshift value.

The improbability of having a {\it single} absorption system to explain 
{\it 
all} the absorption lines in an object exhibiting a redshift around 2.2, 
as discussed 
above, and the fact that the system cannot even identify {\it all} the 
lines in 
the 
observed spectrum, lead to the conclusion that the redshift cannot be 
accepted as the redshift of the host galaxy.\\

3.2. X-ray emission lines\\

The X-ray spectrum exhibits five emission lines which
have been identified to "the closest abundant K$\alpha$ transitions to the
observed lines", based on the redshift z$_{r(op,abs)}$ = 2.14 [9, 10]. The 
mean
redshift is z$_{r(x,em)}$ = z$_{r}$ = redshift of the host galaxy of the
GRB 011211 = 
1.862, computed from these identifications. This does not match at all 
with the
optical absorption redshift, and the difference z$_{r(op,abs)}$ - z$_{r}$  
= 0.278, implying an outflow velocity (v) for the line emitting
material of v/c = 0.085 $\pm$ 0.02 ($\approx$2.5x10$^{4}$ kms$^{-1}$), 
where c is the velocity of light. 
Thus, some "arbitrary blueshifts are invoked to adjust the 
closest atomic transition to match the observed energy of detected 
excesses" [13]. 

Furthermore, the mean redshift of the five emission lines in the X-ray
spectrum is, as mentioned above,
z$_{r}$ = 1.862, but the difference (spread
$\Delta$z$_{r}$) between the maximum redshift of the system
2.03 exhibited by Mg XI and the minimum redshift of the
system 1.73 exhibited by Ar XVIII is $\Delta$z$_{r}$ = 0.3 (see Table
1). The alternative identification of Mg XII instead of MgXI, suggested 
by the
authors [9,10], is actually
more appropriate for a Ly$\alpha$-like transition, but then,
the redshift is 2.3409 and $\Delta$z$_{r}$ = 0.6109. No reason 
has been given by these authors for {\it not} adopting this alternative 
identification.

Ideally, spread values should be close to zero. However, upto a certain 
extent, spreads can have some physical reasons, mainly
because of the difficulty encountered in the exact determination of the
observed wavelength. The
latter, in its turn, may have reasons of its own, viz. the profile being
broad or double- or multi-peaked or of complex structure, blending,
intrinsic
or intervening absorption, gradients, net flows, partial screening, etc. 
The unacceptably high value $\Delta$z$_{r}$ = 0.3, let alone
$\Delta$z$_{r}$ = 0.6109, cannot be explained by any physical mechanism.\\

{\bf 4. The alternative blueshift determination in GRB 011211\\}

The blueshift (z$_{b}$) of an object is determined by the relation z$_{b}$ 
= ($\lambda_{e}$-$\lambda_{o}$)/$\lambda_{e}$, where $\lambda_{e}$ and 
$\lambda_{o}$ are emitted and observed wavelengths respectively in 
\AA. $\lambda_{e}$ is obtained from the search list of known 
laboratory lines. An extended search list covering the UV, optical and 
IR regions has been prepared and is available in [30]. $\lambda_{o}$ is 
obtained from the record of 
observation. In the X-ray region, where wavelengths are usually expressed 
in energy units, the blueshift is determined by z$_{b}$ = 
(E$_{o}$-E$_{e}$)/E$_{o}$, where E$_{e}$ and E$_{o}$ are emitted and 
observed values of wavelengths respectively in keV. E$_{e}$ is available 
in standard tabls and E$_{o}$ is obtained from the observed spectra.

We have interpreted both the optical absorption spectrum (current superior 
quality data of Vreeswijk et al. [8]) and the X-ray 
emission spectrum, in terms 
of blueshifts. Table 1 shows the identifications of the observed lines 
in the spectra of GRB 011211 in both redshift and blueshift 
interpretations. We 
have followed the standard procedure in the identification process 
[46,47], viz. a 'shift' (red or blue) is only confirmed  when 
at least two observed lines, emission spectrum or absorption spectrum, 
exhibit the same 
value when identified with two separate search lines. Any third or more 
lines seen in the spectrum, and, in case of absorption lines, belonging to 
the 
same system, have also to obey the same value. Further, 
in our identification, if and when, the lower order line(s) of a series 
and/or the stronger component of a doublet are/is identified, the higher 
order line(s) of the series and/or the weaker component of the doublet may 
be too weak to be seen. If and when, however, the higher order 
line(s) 
of a series and/or the weaker component of a doublet are/is identified, 
the lower 
order line(s) of the series and/or the stronger component of the doublet 
are/is outside 
the observed region of the spectrum.

For the optical spectra, {\it all} the observed lines have been 
identified with alternative search lines of longer wavelengths, 
including the line at 6114.2\AA which the redshift hypothesis failed to 
identify. In the blueshift identification, the seventeen
absorption lines seen in the optical spectrum are absorbed in six
separate clouds representing six separate systems,
viz. (i) H$\alpha$ 6563, HeI 7065, OI 8449 (ii) OI 8449, HeII 10124, OI
11210 (iii) OI 8449, HeII 10124, P$\beta$ 12818 (iv) OI 11210,  
P$\beta$ 12818, P$\alpha$ 18751 (v) H2 19750, H2 21218, H2 21542 (vi) HeI 
17008, P$\alpha$ 18751. 

It may be noted that all these are well recognized search lines used 
regularly in redshift identification programs as well. Further, 
all the blueshifted lines are permitted lines to match the observed broad 
lines and chosen also to match the strengths of the observed lines as 
given by the equivalent widths. The line at 3820\AA identified in the 
redshift system as 
Ly$\alpha$ yields the rest frame equivalent width of 5.28\AA which is too 
small as it is known as one of the strongest search lines. The blueshift 
identification, on the other hand, yields the H$\alpha$ rest frame 
equivalent width 
of 28.52 which matches the strength of the observed line. Same is true 
for the lines at 
4863.4\AA and 4870.9\AA which have been identified in the redshift 
scenario 
with CIV, another of the strongest search lines in the redshift scenario, 
yielding rest frame 
equivalent widths of 0.7\AA and 1.18\AA respectively, which are again too 
small. The blueshift 
identifications of these lines with medium strong search lines OI 8449 and 
HeII 10124 [30] yielding rest 
frame equivalent widths of 3.83 and 7.69 respectively are, once again 
better fits.

Furthermore, the line at 6114.2 remains unidentified in the redshift 
system, 
and the blueshift system has duly identified the line fitting it with 
other 
lines of one of the systems.

Again, SiIV 1393/1402 
is a medium strong line [30], and its equivalent width 0f 0.92/0.64 is 
much smaller than expected. These lines at 4318.7/4405.0 identified 
respectively with 
OI 8449 
and OI 11210 yielding rest frame equivalent widths of 5.59/5.09 are 
certainly better matches.

It is worth furher noting that redshifts and blueshifts are determined by 
identifying the principal line(s) first and other lines then follow 
matching 
the redshift or blueshift value(s) thus obtained. As pointed out above, 
the 
redshift hypothesis has failed to identify properly the most prominent 
feature in the absorption spectrum, viz. the line at 3829.0\AA.

On the other hand, the six absorption systems have mean 
blueshift values
z$_{b(op,abs)}$ 0.4208,
0.4809, 0.5249, 0.6042, 0.6205, 0.7487, with corresponding spreads, 
$\Delta$z$_{b(op,abs)}$, calculated as the 
difference between the maximum and minimum blueshift values of each 
system, 0.0065, 0.0013, 0.0139,
0.0062, 0.0070, 0.0046 respectively. Unfortunately, uncertainties in the  
determination of the observed wavelengths ($\lambda_{o}$) for the optical 
absorption lines are not available. But published 
literature would confirm that, usually, for absorption lines in 
extragalactic 
objects, $\Delta$z$_{r(op,abs)} <$ 0.01. With one exception, viz. the 
system z$_{b(op,abs)}$ = 0.5249, with $\Delta$z$_{b(op,abs)}$ = 
0.0139, all values are within this limit. Such exceptions are, of course, 
not 
unheard of in the redshift literature. The optical absorption redshift 
system 2.8102 in the QSO 0528-250 has a spread of 0.0173 [48,49]. 

The five emission lines in the X-ray spectrum are identified with
K$\alpha$ transitions of Ne,
O, N and C, and L$\alpha$ transition of Fe, and the blueshift of each 
line is determined. The uncertainty in the 
determination of the blueshift of each X-ray line is also shown in Table 
1, based on the uncertainty in the determination of each observed 
wavelength. The identifications yield  
z$_{b(x,em)}$ = z$_{b}$ = 0.4115, the mean blueshift of the host galaxy of
the GRB 011211, and $\Delta$z$_{b}$ = 
0.0768$\pm$0.073. This spread is consistent with $\Delta$z$_{b}$ = 0 at 
the confidence level of $\sigma$ = 1.05. 

Also, the six elements, viz. Ne, O, N. C, H and Fe 
are among the most abundant elements in the universe, and are often 
identified in extragalactic objects, and the K$\alpha$ transition is the 
strongest transition in X-ray spectra.

Moreover, the 
detection of Fe L$\alpha$ has
been reported earlier [50] in galaxies M 82 and NGC
253, where Fe
K$\alpha$ is very weak (doubtful) or not seen at all. 
At the blueshift 0.4174 for the Fe L$\alpha$, Fe K$\alpha$ is expected 
$\approx$10.1 keV, which is outside the observed
wavelength range of Reeves et al [9,10]. 

The six optical absorption
systems with blueshifts larger than that of the host galaxy are located in
the intervening space along the line of sight. As discussed earlier (Sec. 
3.1),
this is the expected scenario rather than {\it all} the lines being
absorbed in the host galaxy, and no absorbing cloud encountered in the
intervening path. We propose that the
host galaxy, along with the six associated absorbing clouds, have
been ejected, as described in the following scenario (Sec. 5).\\ 

{\bf 5. A generic proposal: ejection mechanism\\}

It is known that supermassive black holes are seats of activities at 
centres of galaxies [51,52]. When the 
system becomes gravitationally unstable due to 
strong interactions at the centre, one or more massive objects may be 
ejected by the so called "sling-shot" mechanism [53-55]. The scenario has 
been further developed as 
follows. 

Two galaxies, each hosting a supermassive black hole may merge 
resulting in the initial formation of a binary system containing the two 
central black holes [56]. Binary black hole systems have indeed 
been detected at X-ray wavelengths in NGC 6240 [57], possibly in OJ 287 
[58], and very recently in SDSS J153636.22+044127.0 [59]. 
As the merger process proceeds further, a single black hole is ejected at 
a relativistic or non-relativistic speed, if the two individual black 
holes are of unequal masses [60]. Evidence of 
ejection of a supermassive black hole by the "sling shot" mechanism 
resulting from merger of galaxies has recently been presented by Haehnelt 
et al. [61]. Again, it is 
believed [62-64] that 
the black hole seated at the centre of a galaxy is often surrounded by a 
gaseous accretion disk which survives the tidal disruption involved in the 
ejection process. Several authors have shown that the interaction between 
the surroundings and the disk associated with the black hole may be 
responsible for the production of galaxy-like objects [65-67].

It is also known that the central supermassive black holes 
(primaries) may be accompanied by satellite black holes of intermediate 
masses [68], and a "small black hole 
swarm around the supermassive black hole in the core of the Milky Way" has 
recently been reported by Munro (AAS meeting January 2005, Sky \& 
Telescope 
April 2005). Satellite black holes are ejected too as a result of the 
merger process, and at least some of them may assume eccentric orbits 
around the primary [67].

It is reasonable to assume that satellite black holes are also surrounded 
by similar gaseous disks, and would undergo similar interactions with 
their 
surroundings, as in primaries, although at reduced scales, being of 
smaller masses, and would end up as faint or nascent or smaller galaxies. 
The final result of the merger of two galaxies is, therefore, the ejection 
of a new galaxy, along with several galaxy-like 
objects. The latter acts as absorbing clouds when falling along the line 
of 
sight, and, being ejected at larger speeds, exhibit larger blueshifts. It 
may be noted in this connection that Basu [38] had shown earlier 
that the appearance of observing clouds in an extragalactic object 
may be associated with the creation of the object itself. Moreover, 
ejection mechanism 
is well known in the literature, and observations [69-71] support such
systems,
viz. galaxies associted with possible absorbers, the latter being in forms
of other galaxies, faint and nascent galaxies.\\

{\bf 5. Concluding remarks\\}

All extragalactic objects do not exhibit blueshifted spectra. As
such, blueshifts do not contradict redshifts, but complement them. Modern
observational technology is leading to the discovery of larger number of
objects and it appears that some spectra cannot be interpreted as
redshifted. While several researchers suggested that
blueshifts are possible, observed spectra are routinely interpreted in
terms of redshifts only. Considering its possible impact on modern
cosmology, possibility of blueshifts should be included in current line
identification programs.

{\bf Acknowledgement:} The author is grateful to an anonymous referee 
for making helpful comments and suggestions that led to major 
improvement of the paper.

\pagebreak

\def\baselinestretch{1.00}
\large
\normalsize

\begin{tabular}{lllllllllr}
\multicolumn{10}{c}{{\bf Table 1} Redshifts and blueshifts in the 
spectra of the GRB 011211}\\ \hline
\multicolumn{1}{c}{TYPE}
& \multicolumn{1}{c}{$\lambda_{o}$}
& \multicolumn{1}{c}{W$_{o}$}
& \multicolumn{1}{c}{z$_{r}$ line}
& \multicolumn{1}{c}{z$_{r}$}
& \multicolumn{1}{c}{W$_{er}$}
& \multicolumn{1}{c}{z$_{b}$ line}
& \multicolumn{1}{c}{z$_{b}$}
& \multicolumn{1}{c}{W$_{eb}$} \\ \hline

ABS & 3820.0 & 16.6 & Ly$\alpha$ 1216 & 2.1423 & 5.28 & H$\alpha$ 6563 &
0.4179 & 28.52 \\
    & 3957.0 & 4.7 & SiII 1260 & 2.1394 & 1.50 & OI 8449 & 0.5328 & 10.06 
\\
    & & & SiII 1259 \\ 
    & 4096.9 & 4.6 & SiII 1304 & 2.1419 & 1.46 & HeII 7065 & 0.4201 
& 7.93 \\
    &  & & OI 1302 \\
    & 4195.6 & 5.6 & CII 1334 & 2.1439 & 1.78 & HeI 17008 & 0.7533 &
22.7 \\
    & & & CII 1335 \\
    & 4318.7 & 2.9 & SiIV 1393 & 2.1438 & 0.92 & OI 8449 & 0.4814 & 5.59 
\\
    & 4405.9 & 2.0 & SiIV 1402 & 2.1409 & 0.64 & OI 11210 & 0.6070 &
5.09 \\   
    
    & 4797.8 & 3.6 & SiII 1526 & 2.1426 & 1.15 & P$\alpha$ 18751 & 0.7441 
& 14.07 \\
    & 4863.4 & 2.2 & CIV 1548 & 2.1413 & 0.70 & OI 8449 & 0.4244 
& 3.82 \\
    & 4870.9 & 3.7 & CIV 1550 & 2.1410 & 1.18 & HeII 10124 & 0.5189 & 7.69 
\\
    & 5053.7 & 1.7 & FeII 1608 & 2.1420 & 0.54 & P$\beta$ 12818 & 0.6057 & 
4.30 \\
    & 5251.7 & 3.5 & AlIII 1670 & 2.1432 & 1.11 & HeII 10124 & 0.4813 & 
6.75 \\
    & 5828.3 & 1.7 & AlIII 1854 & 2.1424 & 0.54 & OI 11210 & 0.4801 & 3.27 
\\
    & 6114.2 & 1.3 & ? & ? & ? & P$\beta$ 12818 & 0.5230 & 2.73 \\  
    & 7362.8 & 3.1 & FeII 2344 & 2.1408 & 0.99 & H2 19570 &
0.6238 & 8.24 \\ 
    & 7485.5 & 6.5 & FeII 2382 & 2.1415 & 2.07 & P$\alpha$ 18751 & 0.6008 
& 16.28\\
    & 8131.4 & 3.8 & FeII 2586 & 2.1436 & 1.21 & H2 21218 & 0.6168 & 9.92 
\\ 
    & 8171.4 & 3.5 & FeII 2600 & 2.1426 & 1.11 & H2 21542 & 0.6208 & 9.23 
\\ \hline
EM  & 0.44$\pm$0.04 & 0.54 & MgXI 1.35 & 2.03 & 0.18 & CK$\alpha$ 
0.277 & 0.3705$\pm$0.057 & 0.86 \\
    & & 0.59 & MgXII 1.47 & 2.3409 \\
    & 0.71$\pm$0.02 & 1.23 & SiXIV 1.99 & 1.82 & 0.43 & NK$\alpha$ 
0.3924 & 0.4473$\pm$0.016 & 2.22 \\
    & 0.88$\pm$0.01 & 1.41 & SXVI 2.6 & 1.9438 & 0.48 & OK$\alpha$ 
0.5249 & 0.4035$\pm$0.007 & 2.37 \\
    & 1.21$\pm$0.02 & 1.26 & ArXVIII3.3 & 1.73 & 0.46 & FeL$\alpha$ 
0.705 & 0.4174$\pm$0.01 & 2.17 \\
    & 1.46$\pm$0.04 & 1.03 & CaXX 4.07 & 1.79 & 0.36 & NeK$\alpha$0.8486 
& 0.4188$\pm$0.016 & 1.76 \\ \hline     

\end{tabular}

TYPE denotes absorption (ABS) or emission (EM) features. $\lambda_{o}$
and W$_{o}$ are
observed wavelength and observed equivalent width, W$_{er}$ and W$_{eb}$
are emitted equivalent widths based on redshifts and blueshifts, z$_{r}$
line and z$_{b}$ line are search lines ($\lambda_{e}$ or E$_{e}$)
identified in redshift and blueshift scenarios, z$_{r}$
and z$_{b}$ are redshift and blueshift values, respectively.\\  

Top panel, optical spectrum [8],
with $\lambda_{o}$ and all W's in \AA. \\

Bottom panel, X-ray spectrum  with $\lambda_{o}$ and 
all W's in keV. $\lambda_{o}$, z$_{r}$ line, z$_{r}$ from [10], W$_{er}$ 
from [9].       

\pagebreak
    
{\bf References\\}
    
\begin{enumerate}

\item Klebasadel, R. et al. Astrophys. J. {\bf 182}, L85 (1973).
\item Liang, E. et al. Astrophys. J. {\bf 479}, L35 (1997).
\item Paczynski, B. Astrophys. J. {\bf 494}, L45 (1998).
\item Iwamoto, K. et al. Nature {\bf 395}, 672 (1998).
\item Frontera, F. et al. GCN GRB Obs. Rep. No. 1215 (2001).
\item Fruchter, A. et al. GCN GRB Obs. Rep. No. 1200 (2001).
\item Holland, S. et al. Astron. J. {\bf 124}, 639 (2002).
\item Vreeswijk, P. et al. Astron. Astrophys. {\bf 447}, 145 (2006).
\item Reeves, J. et al. Nature {\bf 416}, 512 (2002).
\item Reeves, J. et al. Astron. Astrophys. {\bf 403}, 463 (2003).
\item Borozdin, K. N and Trudolyobov, S. P. Astrophys. J. {\bf 583}, L57 
(2003).
\item Rutledge, R. E. and Sako, M. Mon. Not. Roy. astyron. Soc. {\bf 339}, 
600.
\item Sako, M. et al. Astrophys. J. {\bf 623}, 973 (2005).
\item Butler, N. et al. Astrophys. J. {\bf 627}, L9 (2005).
\item Putsil'nik, S. Astron. Astrophys. {\bf 78}, 284 (1979).
\item Gordon, K. Amer. J. Phys. {\bf 48}, 524 (1980).
\item Popowski, P. and Weinzriel, W. Mon. Not. Roy. Astron. Soc. {\bf 
348}, 235 (2004).
\item Weymann, R, et al. Ann. Rev. Astron. Astrophys. {\bf 19}, 41 (1981).
\item La Fevre, O. et al. Astrophys. J. {\bf 471}, L11 (1996).
\item Pascarelle, S. et al. Nature {\bf 383}, 45 (1996).
\item Basu, D. Astrophys. Space Sci. {\bf 259}, 415 (1998).
\item Chen, H. W. et al. Nature {\bf 398}, 586 (1999).
\item Chen, H. W. et al. Nature {\bf 408}, 562 (2000).
\item Stern, D. et al. Nature {\bf 408}. 560 {2000}.
\item Basu, D. Astrophys. Letts. Commun. {\bf 40}, 157 (2001).
\item Wampler, E. and Oke, J. Astrophys. J. {\bf 148}, 695 (1967).
\item McDowell, J. et al. Astrophys. J. {\bf 450}, 585 (1995).
\item Jannuzi, B. Astrophys. J. Suppl. {\bf 118}, 1 (1998).
\item Basu, D. Phys. Scr. {\bf 69}, 427 (2004).
\item Basu, D. and Haque-Copilah, S. Phys. Scr. {\bf 63}, 425 (2001).
\item Basu, D. Astron. J. {\bf 131}, 1231 (2006).
\item Basu, D. J. Astrophys. Astron. {\bf 27}, 381 (2006).
\item Kulkarni, S. et al. Nature {\bf 393}, 35 (1998).
\item Basu, D. Astrophys. Letts. Commun. {\bf 40}, 225 (2001).
\item Riess, A. et al. Astron. J. {\bf 116}, 1009 (1998).
\item Basu, D. Mod. Phys. Letts. A {\bf 15}, 2357 (2000).
\item Basu, D. Astron. Nachr. {\bf 327}, 724 (2006).
\item Basu, D. Astrophys. Letts. {\bf 22}, 139 (1982).
\item Moller, P. et al. Astron. Astrophys. {\bf 296}, L21 (2002).
\item Mirabell, N. et al. Astrophys. J. {\bf 595}, 935 (2003).
\item Mirabell, N. et al. Astrophys. J. {\bf 587}, 128 (2003).
\item Berger, E. et al. Astrophys. J. {\bf 587}, L5 (2003).
\item Masetti, N. et al. Astron. Astrophys. {\bf 465}, 481 (2003).
\item Masetti, N.et al. Astron. Astrophys. {\bf 374}, 382 (2001).
\item Metzger, M. Nature {\bf 387}, 878 (1997).
\item Basu, D. Nat. Phys. Sci. {\bf 241}, 159 (1973).
\item Basu, D. The Observatory {\bf 93}, 229 (1973).
\item Smith, H. et al. Astrophys. J. {\bf 228}, 369 (1979).
\item Mayor, D. and York, D. Astrophys. J. {\bf 319}, L45 (1987).
\item Ptak, A. et al. Astron. J. {\bf 113}, 1286 (1997).
\item Basu, D. et al. Astron. Astrophys. {\bf 272}, 417 (1993).
\item Capetti, A. et al. Astron. Astrophys. {\bf 431}, 465 (2005).
\item Saslaw, W. et al. Astrophys. J. {\bf 190}, 253 (1974).
\item Valtonen, M. Astron. Astrophys. {\bf 46}, 429 (1976).
\item Valtonen, M. Astron. Astrophys. {\bf 46}, 435 (1976).
\item Valtaoja, L. et al. Astrophys. J. {\bf 343}, 47 (1989).
\item Komossa, S. at al. Astrophys. J. {\bf 582}, L15 (2003).
\item Valtonen, M. et al. Astrophys. J. {\bf 643}, L9 (2006).
\item Boronson, T. and Lauer, T. Nature {\bf 458}, 53 (2009). 
\item Mikkola, S. and Valtonen, M. Astrophys. J. {\bf 348}, 412 (1990).
\item Haehnelt, M. et al. Mon. Not. Roy. Astron. Soc. {\bf 366}, L22 
(2006). 
\item Rees, M. and Saslaw, W. Mon. Not. Roy. astron. Soc. {\bf 171}, 53 
(1975).
\item Lin, D. and Saslaw, W. Astrophys. J. {\bf 217}, 958 (1977).
\item De Young, D. Astrophys. J. {\bf 211}, 329 (1977).
\item Rees, M. Ann. Rev. Astron. Astrophys. {\bf 22}, 471 (1984).
\item Osterbrock, D. and Mathews, W. Ann. Rev. Astron. Astrophys. {\bf 
24}, 171 (1986).
\item Valtonen, M. and Basu, D. J. Astrophys. Astron. {\bf 12}, 91 (1991).
\item Carr, B. Comm. Astrophys. {\bf 7}, 161 (1978).
\item Dressler, A., et al. Astrophys. J. {\bf 404}, L45 (1993).
\item Francis, P., et al. Astrophys. J. {\bf 457}, 490 (1996).
\item Giavalisco M., et al. Astrophys. J. {\bf 425}, L5 (1994).

\end{enumerate}

\end{document}